\documentclass[12pt]{iopart}
\usepackage{epsf}
\usepackage{iopams}
\begin{document}

\title
[Metastable bound state of a pair of two-dimensional spatially
separated electrons] {Metastable bound state of a pair of
two-dimensional spatially separated electrons in anti-parallel
magnetic fields}

\author{S. I. Shevchenko
\footnote[1]{shevchenko@ilt.kharkov.ua} and E. D. Vol }

\address{B. I. Verkin Institute for Low Temperature Physics and
Engineering National Academy of Sciences of Ukraine, Lenin av. 47
Kharkov 61103, Ukraine}

\begin{abstract}
We propose a new mechanism for binding of two equally charged
carriers in a double-layer system subjected by a magnetic field of
a special form. A field configuration for which the magnetic
fields in adjacent layers are equal in magnitude and opposite in
direction is considered. In such a field an additional integral of
motion - the momentum of the pair $\vec{\cal P}$ arises. For the
case when in one layer the carrier is in the zero ($n=0$) Landau
level while in the other layer - in the first ($n=1$) Landau level
the dependence of the energy of the pair on its momentum $E({\cal
P})$ is found. This dependence turns out to be nonmonotonic one :
a local maximum and a local minimum appears, indicating the
emergence of a metastable bound state of two carrier with the same
sign of electrical charge.
\end{abstract}




\section{Introduction}
During last ten years a possibility to measure the effects caused
by an interaction of spatially separated carriers in
low-dimensional systems has  been demonstrated in a number of
experiments. An undoubted evidence for such effects was obtained
in drag experiments in which a voltage in one conducting layer
caused by an electric current in the adjacent layer (separated
from the first one by a dielectric layer) was observed. The drag
effects have been registered in bilayer systems with the
conductivity of the same type in both layers (for instance, the
electron-types)\cite{1,2,3} and in the layers with the
conductivity of the opposite types (electron-type in one layer and
the hole-type in the other one) \cite{4,5}. In the last case the
interaction between spatially separated electrons and holes may
result not only in the drag effect but also in an electron-hole
pairing. The electron-hole pairs may condense into a specific
superfluid state in which a supercurrent in one layer is
accompanied by a supercurrent in the adjacent layer and these
currents have the same absolute value but the opposite directions
\cite{6,7}.

The most favorable conditions for the electron-hole pairing are
achieved in a case when a strong (quantizing) perpendicular to the
layers magnetic field is applied to a bilayer electron-hole
\cite{8,9,10} or electron-electron system \cite{11,12} (with the
total filling factor $\nu_T=1$ in the last case). The experimental
discovery of the superfluidity of the pairs in such systems has
been already reported \cite{13,14}.

The possibility of pairing of spatially separated electrons and
holes looks quite natural since there are Coulomb attraction
forces between an electron and a hole. Unexpected and less obvious
phenomenon consists in that in a strong  magnetic field the
Coulomb repulsion between spatially separated equally charged
carriers may  result in a formation of metastable bound pairs.
Such an effect takes place in a situation when the magnetic fields
applied to the first and the second layer of bilayer electron (or
hole) system are antiparallel to each other. This possibility was
predicted in our paper \cite{15}, where we assume that together
with the antiparallel perpendicular to the layers magnetic fields
the antiparallel to each other and parallel to the conducting
layers electric fields are applied to the system. The disadvantage
of the situation considered in \cite{15} is that the presence of
the electric fields may result in an instability of the system
with a macroscopic number of the pairs.

In this paper we show that a formation of a metastable bound state
of spatially separated electrons (or holes) can emerge without the
electric fields applied to the system and formulate the conditions
of appearance of such a bound state.

\section{Electron-electron pair in antiparallel magnetic fields}

Let us consider two two-dimensional electron layers with the
interlayer distance $d$  embedded in a dielectric matrix with the
dielectric constant $\varepsilon_0$. Let the magnetic field in the
top layer (layer 1) is ${\bf B}_1=(0,0,-B)$ and the magnetic field
in the bottom layer (layer 2) is ${\bf B}_2=(0,0,B)$ (the $z$ axis
is chosen perpendicular to the layers). The possible way of
realization of such a configuration of magnetic fields will be
discussed in the end of the paper. We specify the case when there
is one electron belonging to the zero Landau level in the layer 1
 and one electron belonging to the first Landau level in the
layer 2. In the symmetric gauge the vector potential in the layer
1 is equal to ${\bf A}_1({\bf r}_1)=(B y/2, -B x/2,0)$ and in the
layer 2 ${\bf A}_2({\bf r}_2)=(-B y/2, B x/2,0)$. The Hamiltonian
of a pair of interacting electrons has the form
\begin{equation}\label{1}
  H=H_1+H_2+V_C(|{\bf r}_1-{\bf r}_2|),
  \end{equation}
where
\begin{equation}\label{2}
  H_1=\frac{\left(\hat{p}_{1x}+\frac{eB}{2c} y_1\right)^2}{2
  m_1}+\frac{\left(\hat{p}_{1y}-\frac{eB}{2c} x_1\right)^2}{2 m_1},
\end{equation}
\begin{equation}\label{3}
  H_2=\frac{\left(\hat{p}_{2x}-\frac{eB}{2c} y_2\right)^2}{2
  m_2}+\frac{\left(\hat{p}_{2y}+\frac{eB}{2c} x_2\right)^2}{2 m_2},
\end{equation}
\begin{equation}\label{4}
  V_C=\frac{e^2}{\varepsilon_0|{\bf r}_1-{\bf r}_2|}=
  \frac{e^2}{\varepsilon_0 \sqrt{(x_1-x_2)^2+(y_1-y_2)^2+d^2}}.
\end{equation}
Here ${\bf r}_1$, ${\bf r}_2$ are the two-dimensional vectors. We
set the electron charge equals to $-e$.

In a strong magnetic field the Bohr radiuses of the electrons
$a_B^{(1)}=\varepsilon_0 \hbar^2/m_1 e^2$,
$a_B^{(2)}=\varepsilon_0 \hbar^2/m_2 e^2$ can be much larger then
the magnetic length $\ell_0=(c\hbar/eB)^{1/2}$. In this case the
Coulomb interaction can be taken into account as a perturbation.
It is known that the quantum problem of a particle in a quantizing
magnetic field has a large degeneracy (in the symmetric gauge with
respect to the quantum number $m$, the $z$ projection of the
angular momentum). Therefore, the common  formulation of the
theory of perturbations should be based on a solution of a secular
equation. But such an approach is not an optimal method for the
study of this problem. Here we use another approach based on the
projection of the Hamiltonian (\ref{1}) into the subspace of the
states of the pair of electrons in which the electrons in the
layer 1 and 2 are frozen on the zero  and the first Landau levels,
correspondingly (it is just the approach used in the theory of the
quantum Hall effect). Then the kinetic energy operator for the
electron in the layer 1
\begin{equation}\label{5}
  H_1=\frac{(\Pi_{1x})^2+(\Pi_{1y})^2}{2
  m_1}=\hbar\omega_1(a_1^+a_1+\frac{1}{2})
\end{equation}
is projected to $\bar{H}_1=\hbar\omega_1/2$ (here and further the
bar symbols indicate the projected operators) which is constant.
We will omit this constant in the further consideration. In
(\ref{5}) $\omega_1=eB/m_1 c$ is the cyclotron frequency for the
electron in the layer 1, $\Pi_{1x}\equiv p_{1x}+\frac{e B}{2 c}
y_1$, $\Pi_{1y}\equiv p_{1y}-\frac{e B}{2 c} x_1$, the electron
kinematic momentum components,
$a_1^+=\frac{\ell_0}{\hbar\sqrt{2}}(\Pi_{1x}-i \Pi_{1y})$,
$a_1=\frac{\ell_0}{\hbar\sqrt{2}}(\Pi_{1x}+i \Pi_{1y})$, the
creation and the annihilation operators for the electron 1 on a
zero Landau level. It follows from the commutation relation for
$\Pi_i^1$ ($[\Pi_{1x},\Pi_{1y}]=i\hbar^2/\ell_0^2$) that
$[a,a^+]=1$ as it should be. Analogously, for the electron in the
layer 2 one can find
\begin{equation}\label{6}
  H_2=\frac{\Pi_{2x}^2+\Pi_{2y}^2}{2
  m_2}=\hbar\omega_2(a_2^+a_2+\frac{1}{2})
\end{equation}
and  $\bar{H}_2=3 \hbar\omega_2/2$.

To project out the $V_C(|{\bf r}_1-{\bf r}_2|)$ operator it is
convenient to rewrite it in a Fourier-representation form (we
follow the procedure \cite{15a})
\begin{equation}\label{7}
  V_C=\frac{e^2}{2\pi\varepsilon_0}\int d^2 k
  \frac{\exp(-k|d|)}{|k|}\exp\left(ik_x(x_1-x_2)+i
  k_y(y_1-y_2)\right),
\end{equation}
where $|k|=\sqrt{k_x^2+k_y^2}$.

The coordinates of the electron in the layer 1 can be presented as
\begin{equation}\label{8}
  x_1=X_1 -\frac{\ell_0^2}{\hbar} \Pi_{1y}, \qquad
   y_1=Y_1 +\frac{\ell_0^2}{\hbar} \Pi_{1x},
\end{equation}
where $X_1$, $Y_1$ are the coordinates of the centers of its orbit
in the magnetic field. The operators $X_1$ and $Y_1$ satisfy the
following commutation relations: $[X_1,Y_1]=-i\ell_0^2$. Besides
that, $X_1$ and $Y_1$ commute with the momenta components
$\Pi_{1x}$ and $\Pi_{1y}$. Analogously, for the electron in the
layer 2 we have
\begin{equation}\label{8a}
  x_2=X_2 +\frac{\ell_0^2}{\hbar} \Pi_{2y}, \qquad
   y_2=Y_2 -\frac{\ell_0^2}{\hbar} \Pi_{2x},
\end{equation}
where $[X_2,Y_2]=i\ell_0^2$. Now the projection of the operator
$V_C$ is reduced to the independent projection of two commuting
operators $U_1$ and $U_2$:
\begin{equation}\label{9}
  \hat{U}_1\equiv\exp\left(-ik_x\frac{\ell_0}{\hbar}\Pi_{1y}
  +ik_y\frac{\ell_0}{\hbar}\Pi_{1x}\right)=\exp\left[\frac{\ell_0}{\sqrt{2}}
  (ka_1^+ -\bar{k}a_1)\right],
\end{equation}
\begin{equation}\label{9a}
  \hat{U}_2\equiv\exp\left(ik_x\frac{\ell_0}{\hbar}\Pi_{2y}
  -ik_y\frac{\ell_0}{\hbar}\Pi_{2x}\right)=\exp\left[\frac{\ell_0}{\sqrt{2}}
  (\bar{k}a_2^+ -k a_2)\right],
\end{equation}
where the notation $k\equiv k_x+ i k_y$ is used. The projection of
these operators can be easily done:
\begin{equation}\label{10}
\fl  \bar{U_1}=\langle 0 | \hat{U}_1 |
  0\rangle=\exp\left(-\frac{|k|^2\ell_0^2}{4}\right),\quad
\bar{U_2}=\langle 1 | \hat{U}_2 |
  1\rangle=\exp\left(-\frac{|k|^2\ell_0^2}{4}\right)
  \left[1-\frac{|k|^2\ell_0^2}{2}\right].
\end{equation}
Substituting Eq. (\ref{10}) into Eq. (\ref{7}) we arrive to the
following expression for the operator $\bar{V}_C$
\begin{equation}\label{11}
  \bar{V}_C=\frac{e^2}{2\pi\varepsilon_0}\int d^2 k
  \frac{e^{-|k|d}}{|k|} e^{-\frac{|k|^2\ell_0^2}{2}}
  \left[1-\frac{|k|^2\ell_0^2}{2}\right]e^{ik_x(X_1-X_2)+i
  k_y(Y_1-Y_2)}.
\end{equation}
Since the operators $X_1-X_2$ and $Y_1-Y_2$ commute with each
other and with the Hamiltonian  (\ref{1}) these operators are the
integrals of motion. The appearance of the integrals of motions in
the problem considered is not accidental. The point is that the
Hamiltonian (\ref{1}) and the Hamiltonian of the electron-hole
pair coincide with each other up to a sign of the Coulomb
interaction. In the last (electron-hole) case there is the
integral of motion - the momentum of the pair $\vec{\cal P}$
\cite{17}
\begin{equation}\label{12}
\vec{\cal P}=\left(-i\hbar \frac{\partial}{\partial {\bf r}_1}
+\frac{e}{c} {\bf A}_1\right)+\left(-i\hbar
\frac{\partial}{\partial {\bf r}_2} -\frac{e}{c} {\bf
A}_2\right)-\frac{e}{c} \left[ {\bf B}\times({\bf r}_1-{\bf
r}_2)\right].
\end{equation}
In our problem the momentum of the pair $\vec{\cal P}$ is also the
integral of motion. Comparing  Eq. (\ref{12}) with Eqs. (\ref{8})
and (\ref{8a}) we find the relation between the components of the
momentum $\vec{\cal P}$ and the operators $X_1-X_2$ and $Y_1-Y_2$:
\begin{equation}\label{13}
\fl  {\cal P}_x =\frac{\hbar}{\ell_0^2}(Y_2-Y_1)\quad {\rm
  and}\quad
 {\cal P}_y =-\frac{\hbar}{\ell_0^2}(X_2-X_1)\quad {\rm
 or}\quad
\vec{{\cal P}} =\frac{\hbar}{\ell_0^2}({\bf R}_2-{\bf R}_1)\times
e_z,
\end{equation}
where $e_z$ is the unit vector in $z$-direction. Taking Eq.
(\ref{13}) into account we rewrite the energy of the electron pair
as the function of its momentum $\vec{\cal P}$:
\begin{equation}\label{14}
  \Delta E_{01}(\vec{\cal P})=\frac{e^2}{2\pi \varepsilon_0}
  \int d^2 k \frac{e^{-|k|d}}{|k|} e^{\frac{ik_x \ell_0^2}{2} {\cal
  P}_y - \frac{ik_y \ell_0^2}{2} {\cal
  P}_x} e^{-\frac{k^2 \ell_0^2}{2}}\left(1-\frac{k^2
  \ell_0^2}{2}\right).
\end{equation}

Prior to analyze Eq. (\ref{14}) we note the following.

1. Using the method presented here one can also find the
dependence of the energy of the pair on its momentum $\vec{\cal
P}$ in a general case, when the electron in the layer 1 is
"frozen" on the $n_1$-th Landau level and the electron in the
layer 2 - on the $n_2$-th level. The final result is
\begin{equation}\label{15}
\fl \Delta E_{n_1 n_2}(\vec{\cal P})=\frac{e^2}{2\pi
\varepsilon_0} \int d^2 k \frac{e^{-|k|d}}{|k|} e^{\frac{ik_x
\ell_0^2}{2} {\cal
  P}_y - \frac{ik_y \ell_0^2}{2} {\cal
  P}_x} e^{-\frac{k^2 \ell_0^2}{2}}L_{n_1}\left(\frac{k^2\ell_0^2}{2}\right)
   L_{n_2}\left(\frac{k^2\ell_0^2}{2}\right),
\end{equation}
where $L_n(x)$ are the Laguerre polynomials.

2. The variant of the theory of perturbations used here allows to
solve the problem also in the case when the condition $a_B\gg
\ell_0$ is satisfied only for the one particle in the pair, while
the Bohr radius for the other particle can be of order $\ell_0$.
Such a situation may take place if the effective masses of the
carries differ considerably from each other ($m_1\ll m_2$), see
\cite{15}.

Returning to the analysis of the result (\ref{14}) we consider
only the simplest case $d\to 0$. The results obtained below remain
qualitatively correct if $d\lesssim \ell_0$. At $d=0$ the integral
in the r.h.s. of Eq.  (\ref{14}) is calculated analytically. The
result is
\begin{equation}\label{16}
\Delta E_{01}(p)=\frac{e^2}{\varepsilon
\ell_0}\sqrt{\frac{\pi}{2}} \frac{1}{2}\left[
\left(1-\frac{p^2}{2}\right){\rm I}_0\left(\frac{p^2}{4}\right)-
\frac{p^2}{2}{\rm I}_1\left(\frac{p^2}{4}\right)\right]
e^{-\frac{p^2}{4}},
\end{equation}
where $p={\cal P}\ell_0/\hbar$ is the dimensionless momentum of
the pair, ${\rm I}_0(x)$ and ${\rm I}_1(x)$ are the modified
Bessel functions of the zero and the first order, correspondingly.
Using the asymptotic expressions for ${\rm I}_0(x)$ and ${\rm
I}_1(x)$ one can find from Eq. (\ref{16}) the dependence $\Delta
E_{01}(p)\equiv\epsilon(p)$  at small and large $p$.

1) At $p\ll 1$
\begin{equation}\label{17}
  \epsilon(p)\cong E_0 +\frac{\hbar^2 p^2}{2 M_* \ell_0},
\end{equation}
where the energy $E_0$ and the effective mass of the pair $M_*$
read as
\begin{equation}\label{18}
  E_0=\frac{e^2}{2\varepsilon_0 \ell_0} \sqrt{\frac{\pi}{2}},\quad
  {\rm and}\quad M_*=\left(\frac{2}{\pi}\right)^{1/2}\frac{4
  \varepsilon_0 \hbar^2}{e^2 \ell_0}.
\end{equation}
One should note that in the approximation used the effective mass
$M_*$ is determined only by the interaction between electrons. The
bare masses $m_1$ and $m_2$ do not enter into the expression for
$M_*$. The bare masses determine only the Larmour frequencies
$\omega_1$ and $\omega_2$.

2) At $p\gg 1$
\begin{equation}\label{19}
  \epsilon(p)\simeq \frac{4}{\sqrt{2\pi}}\frac{E_0}{p}.
\end{equation}
As it follows from Eq. (\ref{18}) and (\ref{19}) the energy of the
electron pair as the function of the momentum $p$ increases at
small momenta and decreases at large $p$. Numerical estimates show
that the function $\epsilon(p)$ reaches its maximum
$\epsilon_m=1.148 E_0$ at $p=p_m=1.194$. The dependence
$\epsilon(p)$ is shown in Fig.1.

\begin{figure}
\begin{center}
\epsfbox{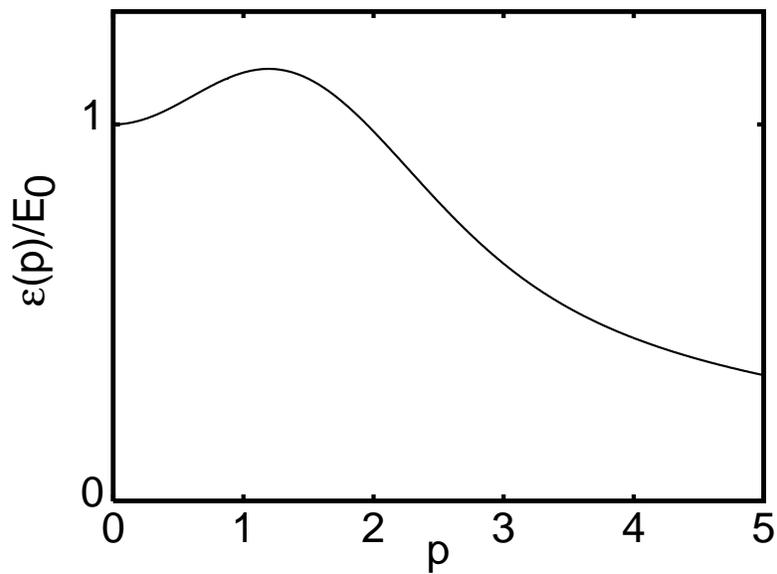}
\end{center}
\caption{\label{f1}The dependence of the energy of the electron
pair on its momentum.}
\end{figure}

\section{Discussion}
The existence of the local minimum of $\epsilon(p)$ at $p=0$ is
the main result. It has important consequences. There are simple
and transparent reasons for the appearance of the local minimum in
the case when the electron in one layer belongs to the zero Landau
level and the electron in the adjacent layer - to the first Landau
level.  The wave functions are shown schematically in Fig. 2
(rigorously speaking,  the electron density distributions in the
layers are presented in this figure).

\begin{figure}
\begin{center}
\epsfbox{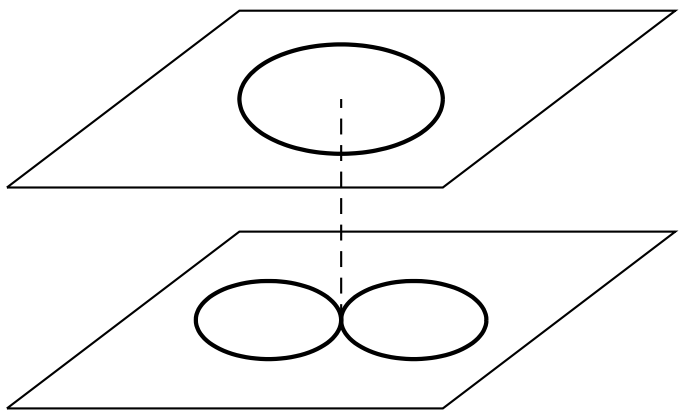}
\end{center}
\caption{\label{f2}Configuration of the electron density in two
layers in the $p=0$ pair state.}
\end{figure}

It is physically evident that in the case when the centers of the
orbits of both electron coincide ($p=0$) the energy of the Coulomb
repulsion is smaller (local minimum) than at small displacements
from this configuration. One should note that the interaction
energy $\epsilon(p)$ is positive at all $p$, which is natural
since the repulsion forces act between the electrons. But the
presence of the minimum gives an evidence of a formation of a
metastable bound state of two electrons from the adjacent layers.
Actually, it is easy to find the average distance between the
electron of the pair with the momentum $\vec{\cal P}$
\begin{equation}\label{20}
  r\equiv
  \left(\langle\psi|(r_1-r_2)^2|\psi\rangle\right)^{1/2}=\ell_0(4+p^2),
\end{equation}
where $|\psi\rangle$ is the wave function of the pair with the
momentum $\vec{\cal P}$. Therefore, these is a one to one
correspondence between the momentum and the size of the pair. If
the pair has the momentum ${\bf p}=0$ then the average distance is
$r=2\ell_0$. To separate the electrons on the infinite distance
from each other (when $\epsilon(p=\infty)=0$) the energy barrier
$\epsilon(p_m)-\epsilon(0)=0.148 E_0$ should be overcome. It means
that the electron pairs with small momenta $p$ are in the
quasibound state  stable against different scattering processes in
which the shift of the momentum of the pair is smaller then $p_m$.
The bound electron pairs are the bosons and, therefore, in the
system of such pairs one can expect the transition into the
superfluid state if the density of the pair is quite large. Since
the pairs are charged this state should be superconductive one.
Strictly speaking, the problem of a transition of the electron
pairs into the superfluid state in the system studied requires
further analysis. Since the bound state of the pairs does not
correspond to the true minimum of the energy the thermodynamic
arguments cannot be used to establish the conditions of the
existence of the superfluid state. To clarify this question the
probability of the transition from the state with a given number
of the pair with zero or small momenta should be found.  We expect
that due to the presence of the barrier for pair decoupling in the
two-particle problem a barrier for the destruction of the coherent
state will exist in the many-particle problem as well. Therefore,
at the temperature smaller then the value of the barrier the time
of life of the coherent state will be large.

In conclusion, we discuss shortly the question about the
possibility to realize the magnetic field configuration required.
This question is not so simple from the experimental point of
view. Actually, the fields should be of order of $1\div 10$ T. But
we think this obstacle can be overcome. We propose two ways of
possible solution of this problem. First, the required
(antiparallel) configuration of the magnetic fields can be
realized using the magnetized stripes of magnetically-hard
materials (like Dy) deposited on the bilayer structure. In Ref.
\cite{18} such a method was used for designing periodic magnetic
fields with $B_{max}=1$ T with the aim to study the conducting
properties of a two-dimensional electron gas in such fields.
Another possible way to create antiparallel magnetic field
configuration can be based on using antiferromagnetic systems in
which the spins in each layer are ferromagnetically ordered while
they are directed antiparallel in adjacent layers. For example,
such properties demonstrate the layered manganites
(LaSr)$_{n+1}$Mn$_n$O$_{3n+1}$, the compounds widely studied now
(see, for instance, review \cite{19}).

This work is supported by the INTAS grant No 01-2344.

\end{document}